\magnification=\magstep1
\def\bbbr{{\rm I\!R}}

\def\lambdabar{\raise 1.7mm \vbox{\hrule width 2mm}\hskip -2mm $\lambda\ $}

\def\Angstrom{\buildrel\circ\over {\rm A}}

\def\ket#1{{}\mid #1\rangle{}}

\def\obraket#1#2#3{{}\langle#1\mid #2\mid #3\rangle{}}
 \def\dyad#1#2{\mid #1\rangle\langle#2\mid }

\def \s{standard }
\def\ns{nonstandard }
\def\cf{coupling function }
\def\cs{coupling strength }

\def\boxit#1{\vbox{\hrule\hbox{\vrule #1\vrule}}\hrule}

\input ./psbox
\psfordvips

\centerline{\bf Influence of the extent of the eigenstates of a system on the resonances}

\centerline{\bf formed through its coupling to a field}

\medskip

\centerline{Claude Billionnet}

{\it Centre de Physique Th\'eorique, \'Ecole Polytechnique, CNRS, 91128 Palaiseau, France}

\centerline{\sevenrm Electronic address: claude.billionnet@cpht.polytechnique.fr}
\medskip

{\tenrm We examine resonances for two systems consisting of a particle coupled to a massless boson's field. The field is the free field in the whole space. In the first system, the particle is confined inside a ball. We show that besides the usual energy levels of the particle, which have become complex through the coupling to the field, other resonances are to be taken into account if the ball's radius is comparable to the particle's Compton wavelength. In the second system, the particle is in a finite-depth square-well potential. We study the way the resonances' width depends on the extent of the uncoupled particle's wave functions. In both cases, we limit ourselves to considering two levels of the particle only.}

\bigskip
\centerline{\bf I. INTRODUCTION}

\smallskip

In our preceding works [1,2], we introduced two types of resonances, which we called ``standard'' and ``nonstandard'', or ``of ${\cal R}$-type'' and ``of ${\cal C}$-type''. The term ``nonstandard'' is used with an analogous meaning in the expression ``nonstandard multi-quark/gluon states''. ``${\cal C}$'' is related to the fact that it is the continuum which yields nonstandard resonances whereas ``${\cal R}$'' refers to the real eigenvalues to which \s resonances are associated. These two types occur every time a quantum system $S$ with discrete energy-levels is coupled to a massless field. The most well-known of these resonances are of course those which are ``attached'' to the eigenvalues of the Hamiltonian of system $S$ alone. By saying that they are attached to these eigenvalues, we mean precisely that they tend to these eigenvalues when the coupling of the system to the field tends to zero. We called them standard resonances, or of ${\cal R}$-type.

Other resonances exist. We called them nonstandard resonances, or of ${\cal C}$-type. The reason to distinguish them from the others is that they disappear at the limit where the coupling vanishes. Among those, some are also well-known, in certain experimental situations. For example, if an atom is placed inside a cavity with very good mirrors, the cavity being tuned to the transition between the first excited level of the atom and its fundamental level, one sees that the excited level is split into a doublet, the so-called Rabi doublet. Since the cavity is never perfect, the energies of the doublet are in fact complex, i.e. we get two resonances there; one can see that one is standard whereas the other one is nonstandard. This is an argument for considering both types of resonances. But there still exist other resonances besides this doublet (see for instance [2]). All of them are likely to connect to one another, this time for a mathematical reason: in the simplified models we studied, these resonances are the zeros of an explicit multivalued function. In an example [1], making the coupling constant and the width of the continuum seen by the atom vary along a closed real path in $\bbbr^2$ and following the corresponding displacement of a resonance, we showed that its final position may differ from the initial one; in other words we get to another resonance. In particular, a standard resonance may become a nonstandard one or conversely. Therefore one may think that all resonances are to be put on the same footing, at least from a mathematical point of view.

In spite of these two theoretical reasons to consider the two types of resonances, there are many cases in which some of these resonances can be neglected, on account of their imaginary parts being very large compared to those of the others. This is the case for nonstandard resonances occurring when an atom is coupled to the transverse electromagnetic field in the vacuum, as can be seen for the hydrogen atom in a simplified  treatment which only takes two levels into account [2]. On the contrary, besides the above mentioned Rabi doublet, in many cases the imaginary parts of \s and \ns resonances are comparable. It is the case in a rudimentary two-level model for the coupling of a (quark-antiquark)-type system to a massless boson [3].

In a first approach, one may be tempted to say that the importance of the \ns resonances with respect to the others is related to the size of the ratio $\chi$ of the coupling strength to the continuum's width, when the latter is defined which is the case for Lorentzian coupling functions. But it is also clear that $\chi$ being large does not imply that one can forget the standard resonance, since it may become an eigenvalue. Therefore some more thorough investigation has to be performed. The importance of the ratio $\chi$ has long been known. In the above quoted case of the Rabi doublet in a cavity, where the electromagnetic interaction is not strong, the importance of the \ns resonance is indeed due to the width of the continuum seen by the atom being small, as a result of the cavity's quality factor $Q$ being large. Generally, in two-level cases, we could roughly expect some \s and \ns resonances being equally interesting when  $\chi$ is of the order of $1$. However, the position of the resonances is not a function of this ratio, neither even, strictly speaking, a function of these two variables, because of the multivaluedness. A precise analysis has therefore to be performed in each case. For example, we saw examples in which it cannot be decided which one of the \s or \ns  resonance comes near to the reals when $\chi$ becomes large, unless a careful computation is performed: it may not be the same resonance whether it is the coupling which increases or the width which decreases [1].

The role of the continuum's width is seen in the following way, in one dimension, for a system $S$ with two levels $e$ and $f$, when the interaction with the field is given by the simplest Hamiltonian 
$$
\lambda\ \Big( \dyad{\psi_e}{\psi_f}\otimes\  c(g)+\dyad{\psi_f}{\psi_e}\otimes\  c^*(\overline g)\Big),                            \eqno (1)
$$
where $c(g)=\int c_kg(k)dk$, $c_k$ being the annihilator of a boson with wave number $k$. The existence of \ns resonances comes from the fact that the coupling function $g$ becomes infinite, at points at a finite or infinite distance. If $g$ has a pole in the lower half-plane (at a finite distance) and is bounded at infinity, then the limit of the \ns resonances when the coupling goes to zero is $\hbar c$ times this pole. This is why these resonances cannot be seen at zero coupling (since there is no coupling function any more). The pole and the width of $g$ are clearly related in some cases. For example, in the Rabi doublet's case for an atom coupled to the field inside a large-$Q$ cavity, the coupling function may be approximated by a Lorentzian whose width is twice the imaginary part of its pole in the complex plane. The \ns resonance is then interesting because  the pole is close to the reals as a result of the coupling function's width being small. Note that the spreading of the coupling function, even if it makes the determination of a width problematic, does not prevent the \ns resonances from existing.

Let us now note that the poles of $g$ have something to do with the extent in space of the eigenvalues of $H_S$, the Hamiltonian of $S$, in certain types of coupling. This is mentioned in [4]. Let us recall the argument, in one dimension. Assume the field is of the form  $\phi(x)=\sum_k {\cal A}_k\big( e^{-ikx} c_k^*+e^{ikx} c_k\big)$. To $\phi(x)$ we associate the operator $\Phi=\sum_k {\cal A}_k\big( e^{-ikx}\otimes c_k^*+e^{ikx}\otimes c_k\big)$ acting in the product of $L^2(\bbbr)$, the space of states of $S$, and the field state-space,  $e^{\pm ikx}$ acting through multiplication. Consider a coupling $H_I=\lambda\ ({\cal O}\otimes 1)\ \Phi$, where ${\cal O}$ denotes an operator in  $L^2(\bbbr)$. The coupling $-{e\over m}\vec p\cdot\vec A$ of electrodynamics is of this form, but in three dimensions, i.e. $H_I=\lambda(\sum_{n=1,2,3}({\cal O}_n\otimes 1)\Phi_n$. Let us restrict ourselves to the space spanned by $\ket e$ and the $\ket{f,k}$'s, and replace the Hamiltonian of the $S+field$ system by an operator  $H_{\rm red}$ in this space, $H_{\rm red}$ having the same matrix elements as $H$. Setting 
$$
\lambda\ g(k):=\obraket{\psi_f,k}{ H_{\rm red}}{\psi_e}=\lambda\ {\cal A}_k\ \obraket{\psi_f}{{\cal O}e^{-ikx}}{\psi_e},
$$
where the $\psi$'s are here the wave functions in $x$-space, we can write the interaction Hamiltonian in the above form (1). If ${\cal O}$ is a polynomial $P(x,{d\over dx})$, then $\obraket{\psi_f}{{\cal O}e^{-ikx}}{\psi_e}$ has the same analytical properties in $k$ as the Fourier transform of $\overline\psi_f\psi_e$. If $|\overline\psi_f\psi_e(x)|$ is upper bounded at infinity by $Q(x)e^{-|x|/x_0}$, with $x_0>0$ and $Q$ a polynomial, then the Fourier transform is analytic inside the strip  $\{k,|\Im(k)|<x_0^{-1}\}$. If $\overline\psi_f({d\over dx})^n\psi_e(x)$ is exactly of the form  $Q_n(x)\ e^{-|x|/x_0}$, where $Q_n$ is a polynomial, then  $\obraket{\psi_f}{{\cal O}e^{-ikx}}{\psi_e}$ has a pole at $\pm x_0^{-1}\ i$. Thus the pole in the lower half-plane is here directly related to the extent in space (in the exponential sense) of the pair of states $(e,f)$.

The extent of the wave functions, in that sense, is therefore an important parameter as regards the position of the resonances. When the coupling is small, the definition of \ns resonances tells us that the largest the wave-functions' extent is the most interesting the \ns resonances are, since the pole is the closest to the reals. The hydrogen atom is a concrete case on which this relation can be studied. However, in that case, we saw in [2] that considering pairs of very excited states is not sufficient to make these resonances interesting (unless perhaps if time scales much shorter than the life-time of the excited states are considered): the coupling is to small and the poles of the coupling functions are too far from the reals. Besides, the largeness of the imaginary parts is also due to the smallness of the coupling [2]. We are going to resume the study of the influence of that space extent in a strong coupling case, where the \ns resonances are going to be important. Unfortunately, we still will limit ourselves to considering two levels of $S$ only, since the determination of the resonances for more than two levels is a problem too complicated to be dealt with at the moment [5]. When the coupling is strong (note that we do not use here the expression ``strong coupling regime'' which has a special meaning), knowing the poles of $g$ , or its singularities, is not sufficient to get an idea of the position of the resonances. For example, we will see two cases, with coupling functions $g_1$ and $g_2$, where poles of $g_1$ and $g_2$ respectively at $k_1$ and $k_2$ satisfying $k_1<k_2$ yield resonances of widths $\Gamma_1$ and $\Gamma_2$ with $\Gamma_1>\Gamma_2$, whereas the coupling strengths are identical (see pairs (1,2) and (1,3) of Section IV). This might be due to the fact that $g$ has other singularities than its poles: singularities at infinity. The resonances' widths for states of S of various extents should therefore be examined in each particular case.

In Section II, we will recall some results in cases where the coupling function is bounded at infinity, with poles at finite distances. In Section III, we will treat a simple case where the singularities of $g$ are pushed to infinity. This will prepare ourselves for the study of the example of Section IV, for which $g$ has singularities at infinity besides its poles. In this Section III, the extent of the wave functions will be fixed, if the physical parameters of the system are not changed; it will be zero, in the exponential sense. The ratio $\chi$ will therefore be zero. The system which is coupled to the transverse field is a particle inside a ball of radius $a$. Through indicating the role of parameter $a/$\lambdabar$\!\!_C$ ( $\lambda_C$ is the Compton wave-length of the particle) we will show that \ns resonances may be important although $\chi$ vanishes. This will prove that the only consideration of $\chi$ is insufficient to judge the interest of the resonances. Section IV also concerns the strong interaction domain. The particle which is coupled to the field is now put inside a finite-depth square-well. It is a simple system which allows the extent of the system to depend on the excitation level of the particle in the well. Our main aim is to discuss the position of the resonances as a function of this excitation level. It will also be interesting to have the well's geometry vary. It is a mixed case where the coupling function has both poles at finite distances and singularities at infinity.

\bigskip

\centerline{\bf II. EXAMPLES OF RESONANCES WITH A COUPLING FUNCTION}
\centerline{\bf BOUNDED AT INFINITY HAVING POLES AT FINITE DISTANCES}

\medskip

\centerline{\bf A.  A case where a \s  and a \ns resonance come to coincide.}

\smallskip

We first recall two cases for which we can precisely calculate coupling constants and coupling function widths for which a \s and a \ns resonance coincide. In these two examples the above-mentioned ratio $\chi$ is of the order of $1$. The coupling function $g$ in Hamiltonian (1) is Lorentzian in both cases, with a width $\mu$, expressed in ${\cal E}$-units, ${\cal E}$ being the energy difference between the two levels. (We recall that resonances are zeros of an analytic continuation of $f(z)=z-{\cal E}-\lambda^2\int_{-\infty}^\infty g(k)^2(z-\hbar c|k|)^{-1}\ dk$, if the energy of the fundamental is assumed to be $0$; the equation for the eigenvalues of the Hamiltonian is $f(z)=0$.)

In the first case, the (normalized) coupling function is 
$$
\sqrt{2/\pi}\ ({\hbar c\over{\cal E}\mu})^{3/2}{|k|\over 1+({\hbar c k\over{\cal E}\mu})^2}
$$
(see [1]). It has a pole at $k=-i\mu{{\cal E}/(\hbar c})$. Let $\kappa={\cal E}^{-1}\lambda$ denotes the dimensionless coupling constant. Curves of Figures 8 and 9 of [1] indicate that, for $\chi=\kappa/\mu\ll 1$, one expects the width of the \ns resonances which are attached to the pole to be much larger than the \s resonance's one. We show in [1] that some \ns and \s resonances come to coincide for $\kappa=0.5$ and $\mu=1.08$. The ratio $\chi$ (coupling strength/continuum's width) or (coupling strength/pole's imaginary part) is then about $1/2$.

In the second case, the coupling function is now centered  at a value which does not depend on $\mu$. It is 
$$
\sqrt{2/\pi}\ ({\hbar c\over {\cal E}\mu})^{1/2}{1\over 1+\mu^{-2}({c\hbar\over {\cal E}}k-1)^2}\ .
$$
Although this function is neither even nor vanishing at the origin, we used it in [2] since it allows simple calculations in a model approaching the coupling of a two-level atom S with the field in a cavity. The continuum is tuned to the transition in the atom. If $\kappa$ still denotes the coupling strength, in ${\cal E}$-units, some resonances almost coincide for $\mu=0.01$ and $\kappa=0.003$, therefore $\chi\simeq 0.3$.

\medskip
\centerline{\bf B. Review of the hydrogen atom case}

\smallskip

Matrix elements of interaction $-{e\over m}\vec p\cdot\vec A$ of the hydrogen 
atom with the transverse field are no longer Lorentzian functions (see for instance [6]). However they all have poles at finite distances. (Note that these poles may be multiple poles and, since the coupling function occurs through its square, these poles yield twice more resonances than their multiplicities.) The largest the mean extent of the states the closest to the reals the poles are. Indeed, each wave function decreases exponentially with the distance $r$, as $e^{-r/r_0}$ for a certain $r_0$, and, for two levels, one can define a mean extent $\bar r_0$ (harmonic mean of the $r_0$'s of the two levels). When one restrict oneself to a pair of two levels $e_1$ and $e_2$, considering Hamiltonian's matrix elements between states $e_2$ and $(e_1,{\rm 1\  photon})$ only, the coupling function has a (multiple) pole at a complex energy $E=-2 i\  \hbar c\ \bar r_0^{-1}$. The ratio $\chi$ between the coupling strength and the imaginary part of this pole is less than about $10^{-4}$. Therefore we expect \ns resonances to be very far from the reals, in the coupling strength unit for instance; this is confirmed by the calculus [2]. In this example, it is the smallness of the fine structure constant which  entails the smallness of $\chi$. Indeed, on the one hand, the coupling strength, in ${\cal E}$-units, is proportional to  $\sqrt\alpha$ (see formula (20') of [2]). On the other hand, the wave-functions' extent (in the above-considered exponential sense) is small with respect to the transition wavelength (ratio less than $\alpha/2$), that is to say the imaginary part of the pole is large with respect to the energy difference between the two levels. Due to the very large widths of the \ns resonances, the study of the relation of these widths to the extent of the states is a somewhat academic question.

It would of course be interesting to examine the resonances' position when the coupling is strong rather than weak, possibly changing also the mass of the particle. We chose to discuss here the simpler well case (Section IV), even if it has the drawback that the coupling function has singularities at infinity, besides the one at a finite distance. It has also the advantage that the width and the depth of the well are parameters which may be varied.

\bigskip

\centerline{\bf III. THE RESONANCES FOR A PARTICLE CONFINED INSIDE A BALL}

\centerline{\bf AND COUPLED TO A MASSLESS VECTOR FIELD} 

\smallskip

Before considering the well case, we are going in this section to deal with a case where the coupling function is not bounded at infinity but has no pole. 

A particle is confined by a infinite potential barrier, and so as to get simple enough computations with the coupling $-{e\over m}\vec p\cdot\vec A$, we will assume that this barrier is a sphere of radius $a$. The aim is twofold: prepare the study of Section IV, in which $g$ has singularities at infinity, and show how the position of \ns resonances depends on the extension of the states via the ratio of $a$ to \lambdabar$\!\!_C$, $(2\pi)^{-1}$ times the Compton wavelength of the particle. The term ``extension'' is not here taken in the exponential sense but in its ordinary one.

\medskip

\centerline{\bf A. Setting up and physical parameters of the problem}

\smallskip

The states of the system decoupled from the field are those of a particle of mass $m$ in a spherical potential which vanishes for  $r<a$ and is infinite for $r\geq a$. The coupling of the particle to the field is assumed to be given by the interaction Hamiltonian $-\alpha^{1/2}\ (\hbar c)^{1/2}\ m^{-1}\ \vec p\cdot\vec A$. Let us make it clear that the field is not determined by the ball; it is the free field in the whole space. The wave functions  $\Psi_{l,m,p}$ of angular momenta $(l,m)$ are given by the following formulas. Let us set  $\xi_{l,1}\leq\xi_{l,2}\leq\cdots$ the zeros of $j_l(z)=\sqrt{\pi\over 2}\ z^{-1/2}\ J_{l+1/2}(z)$ and
$N_{l,p}:=(\int_0^{\xi_{l,p}} j_l(x)^2 x^2 dx)^{1/2}$.
If $r<a$, $\Psi_{l,m,p}(\vec r)=R_{l,p}(r)\ Y_{lm}(\theta,\phi)$, where $R_{l,p}(r)=N_{l,p}^{-1}\ \xi_{l,p}^{3/2}\ a^{-3/2}\ j_l(\xi_{l,p}r/a)$; if $r>a$, $\Psi_{l,m,p}(\vec r)=0$. The energies of states $(l,m,p)$, independent of $m$, are
$$ 
E_{l,p}=  {1\over 2}({\rm\raise 1.7mm \vbox{\hrule width 2mm}\hskip -2mm}\lambda_C/a) \ (\hbar c/a)\ \xi_{l,p}^2.
$$
As we said before, we forget all the states but two, since we do not know how to calculate otherwise. The upper state, denoted by $e$, has an angular momentum $l_2$ and we assume $m_2=0$. The lower state denoted by $f$ is assumed to have zero angular momentum. Besides, in the coupling of the particle to the field, we only keep matrix elements of the interaction Hamiltonian between  $(f,1\ {\rm boson})$-states and state $e$. In Appendix A we show that the eigenvalues of the particle-field Hamiltonian are the solutions of the equation in $z$
$$
z-E_e-{l_2(l_2+1)\over\pi}({\rm\raise 1.7mm \vbox{\hrule width 2mm}\hskip -2mm}\lambda_C/ a)^2(\hbar c/a)^2\alpha\int_0^\infty{G_{l_2,e,f}(y)^2\over z-E_f-{\hbar c\over a}y}\ {dy\over y}=0,            \eqno (2)
$$
where the (not normalized) coupling function is 
$$
G_{l_2,e,f}(y)=a^4 \int_0^1j_{l_2}(y\rho)\ \rho\  R'_f(\rho a)R_e(\rho a)\ d\rho.\eqno (3)
$$
$y$ is variable $ka$, where $k$ is the boson's wave number.  $G_{l_2,e,f}$ has no dimension; it does not depend on $a$. The interaction strength is measured by the normalized coupling constant
$$
g^{\rm nor}_{e,f}=\sqrt{l_2(l_2+1)\over 2\pi}\ ({\rm\raise 1.7mm \vbox{\hrule width 2mm}\hskip -2mm}\lambda_C/ a)(\hbar c/a)\ ||G_{l_2,e,f}||\ \sqrt\alpha,\quad {\rm with}\quad ||G||=\Big(2\int_0^\infty G(y)^2\ {dy\over y}\Big)^{1/2}.
$$
We call $G^{\rm nor}=||G||^{-1}\ G$ the normalized coupling function. In this example all wave functions have the same extent $a$. Were the extent defined by an exponential-decrease rate, as in the hydrogen atom case or in the finite-depth well case that we will consider in Section IV, we should have to consider this extent as zero. The variable $\mu$ defined in the Lorentzian case or the hydrogen atom case, proportional to the inverse of that extension, would then be infinite. The coupling function's pole is pushed to infinity, becoming an essential singularity, and the ratio  $\chi_{e,f}=\kappa_{e,f}^{\rm nor}/\mu_{e,f}$ vanishes. But here we are going to consider the wave functions' extent as being equal to $a$ and, so as to differentiate the two points of view, the ratio of this extent to $(2\pi)^{-1}$ the wavelength of the considered transition will be denoted by $\nu_{e,f}^{-1}$. We have 
$$
\nu_{e,f}^{-1}={1\over 2}{{\rm\raise 1.7mm \vbox{\hrule width 2mm}\hskip -2mm}\lambda_C\over a}\ (\xi_{l_2,p_2}^2-\xi_{0,p_1}^2).
$$
When it is measured with respect to the transition energy, the coupling strength,\break $\kappa_{e,f}^{\rm nor}:={\cal E}^{-1} g_{e,f}^{\rm nor}$, is
$$
\kappa_{e,f}^{\rm nor}=2\sqrt{l_2(l_2+1)\over\pi}\ \sqrt\alpha\ {1\over \xi_{l_2,p_2}^2-\xi_{0,p_1}^2}\ ||G_{l_2,e,f}||.
$$
Defining $\chi_{e,f}^{\rm dist}:=\kappa_{e,f}^{\rm nor}/\nu_{e,f}$, we get 
$$
\chi_{e,f}^{\rm dist}=\sqrt {l_2(l_2+1)\over\pi}\ \sqrt\alpha\ ||G_{l_2,e,f}||\ {{\rm\raise 1.7mm \vbox{\hrule width 2mm}\hskip -2mm}\lambda_C\over a}.      \eqno (4)
$$
We are going to see that increasing  $\chi_{e,f}^{\rm dist}$ from a value which is small with respect to $1$ to a value of order $1$ change uninteresting \ns resonances into interesting ones. Thus, besides $\chi$, the value of $a/$\lambdabar$\!\!_C$ also plays a role as regards the importance of \ns resonances.

\medskip

\centerline{\bf B. Choice of two states $e$ and $f$}

\smallskip

Through introducing  $\zeta={z-E_f\over E_e-E_f}$, the eigenvalue equation becomes
$$
\zeta-1-l_2(l_2+1)(\kappa^{\rm nor}_{e,f})^2\int_0^\infty{G^{\rm nor}_{l_2,e,f}(y)^2\over \zeta-\nu_{e,f} y}\ {dy\over y}=0.              \eqno (5)
$$
From now on we assume  $l_2=1$ and take the states of lowest energies as $e$ and $f$, i.e we set $p_1=1$ and $p_2=1$. We forget the index $m=0$. The wave functions read:
$$\psi_{0,1}={1\over \sqrt{4\pi}}\ N_0^{-1} \xi_{0,1}^{3/2}\ a^{-3/2}\ {\sin (\xi_{0,1}r/a)\over \xi_{0,1}r/a},
$$
with $\xi_{0,1}=\pi$ and $N_0=\sqrt{\pi/2}\simeq 1.25 $,
$$
\psi_{1,1}= \sqrt{3\over 4\pi}\ N_1^{-1}\ \xi_{1,1}^{3/2}\ a^{-3/2}\ \big({\sin (\xi_{1,1}r/a)\over \xi_{1,1}r/a}-\cos(\xi_{1,1}r/a)\big){1\over \xi_{1,1}r/a} \cos\theta,
$$ 
with $\xi_{1,1}\simeq 4.49$ and $N_1\simeq 1.46$. From (3), the \cf is:
$$\displaylines{
G(y)={(N_0N_1)^{-1}\xi_{0,1}^{3/2}\xi_{1,1}^{1/2}\over y}\times\hfill\cr\hfill\int_0^1\big(({\sin(y\rho)\over y\rho}-\cos(y\rho)\big)\big(\cos(\pi\rho-{\sin(\pi\rho)\over \pi\rho} \big)\big({\sin(\xi_{1,1}\rho)\over \xi_{1,1}\rho}-\cos(\xi_{1,1}\rho) \big)\ \rho^{-2}\ d\rho.  \hfill (6)
}$$ 
Unless we state it otherwise, states ${e,f}$ now remain fixed as was said before and we do not mention them in indices. The norm of $G$, in the above mentioned sense, is $||G||\simeq 2.92$. Since $G$ does not have a more explicit expression than (6), we give the graph of $||G||^{-1}|G|$ in Figure 1.
\noindent
It has to be noted that the width of the highest bump in the curve has nothing to do with the imaginary part of a pole of this function. 

Although we cannot calculate $G$ explicitly, through comparing this function to the coupling function (9) presented in Section IV for the case of a finite depth square well, we expect the former to be unbounded at infinity. Indeed (9) is explicit and its $C_2$-part, which corresponds to the part of the wave function localized inside the well, do have a singularity 
\medskip
\setbox11=\hbox{\psboxto(5cm;0cm){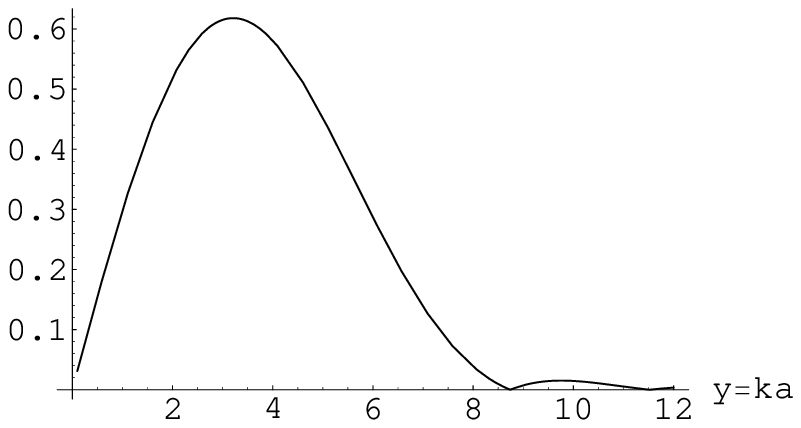}}
\setbox12=\hbox{\psboxto(5cm;0cm){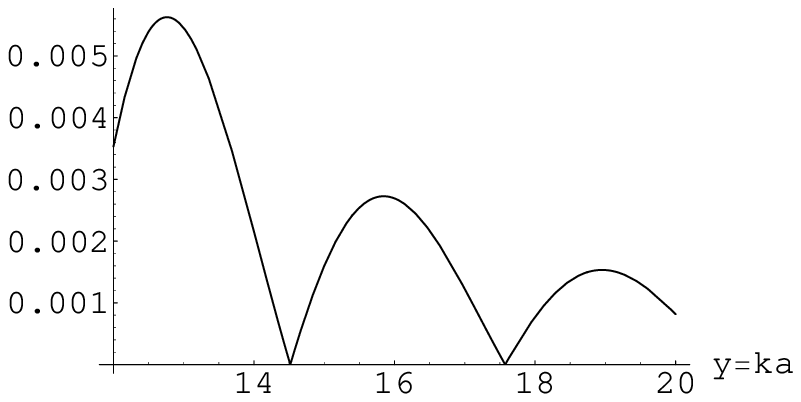}}

\hbox{\hskip 1cm\box11\hskip 2cm\box12}

\vskip 0.2cm
\centerline {FIG. 1. Graph of $||G||^{-1}|G|$} 

\medskip
\noindent
at infinity. We get an analogous oscillation on Figure 1. We recall that this oscillation disappears in the case of the $1/r$ potential; only the pole at a finite distance remains (see Section II.B).

The resonances are given by the equation
$$
\zeta-1-2\ (\kappa^{\rm nor})^2\int_0^\infty{G^{\rm nor}(y)^2\over \zeta-\nu y}\ {dy\over y}+4 i\pi {(\kappa^{\rm nor})^2\over\zeta}G^{\rm nor}(\zeta/\nu)^2=0. \eqno (7)
$$
With the levels we chose, we get
$$
g^{\rm nor}=1.65\ ({\rm\raise 1.7mm \vbox{\hrule width 2mm}\hskip -2mm}\lambda_C/ a )(\hbar c/a)\sqrt{\alpha},\quad \nu\simeq 0.194\ a/{\rm\raise 1.7mm \vbox{\hrule width 2mm}\hskip -2mm}\lambda_C,\quad \kappa^{\rm nor}\simeq 0.32\ \sqrt\alpha,
$$
$$
 \chi^{\rm dist}\simeq 1.65\  ({\rm\raise 1.7mm \vbox{\hrule width 2mm}\hskip -2mm}\lambda_C/ a )\sqrt\alpha.
$$
\medskip

\centerline{\bf C. Numerical application}

\smallskip 

We are going to give two numerical applications. One with the electron, $a=1$ Angstr\"om and $\alpha=1/137$, the other one with a particle whose mass is the reduced mass of two light quarks, i.e. $m=1/2\  m_q=110$ MeV, $a=1$ Fermi and  $\alpha=0.6$. In the first case, \lambdabar$\!\!_C/a=3.86\ 10^{-3}$ and in the second one \lambdabar$\!\!_C/a=1.79$. The results are the following. (We recall that, with the $\zeta$ variable, the levels of the decoupled particle are $\zeta=0$ and $\zeta=1$.) 

\smallskip

\centerline{\bf 1. The electron and an electromagnetic coupling}

\smallskip

If $a$ is expressed in  Angstr\"om, we get $E_{l,p}=0.985\  $(\lambdabar$\!\!_C/a)\ \xi_{l,p}^2$ KeV. Besides,\break $\chi^{\rm dist}=0.14$ \lambdabar$\!\!_C/a$. With  $a=1\  \Angstrom$, we get
$$
E_{0,1}=37.5\  {\rm eV},\quad E_{1,1}=76.7\  {\rm eV},
$$
$$
g^{\rm nor}\simeq 1.07\ {\rm eV},\quad \kappa^{\rm nor}=2.74\ 10^{-2},\quad\nu=50.35,\quad \chi^{\rm dist}=5.44\ 10^{-4}.
$$

One resonance is given by 
$$
\zeta= 53-778\ i, \quad\quad {\rm which\  gives} \ \ z\simeq (92-30\ i)\ {\rm KeV}.
$$
This resonance is thus of no interest. Here $a/$\lambdabar$\!\!_C\simeq 260$.

However, since at the Fermi and GeV scales of Section III.C.2 we will encounter an analogous resonance, no longer negligible, let us give some precisions. Its origin is not unknown of us. It can be shown numerically that if we make $\nu$ tend to $0$ in (7), for example through making $a$ tend to $0$, then this resonance becomes an eigenvalue (solution of (5)) when $\nu$ gets smaller than a certain critical value $\nu_c$ ($\simeq 4.18\ 10^{-4}$, depending on $(e,f)$). For $\nu=\nu_c$, the eigenvalue is zero: the resonance passes below the fundamental $E_{0,1}$. We already met and explained such a situation in Section 3.2 of [4]: for $\nu=0$, the coupling reduces to the coupling of two states of the particle-field system and it is easy to calculate the two eigenvalues $\zeta_+\simeq 1.006$ and $\zeta_-\simeq -0.006$ of the total Hamiltonian. It is toward the latter that the resonance we consider goes if $\nu$ goes to $0$. Besides, we check that this resonance is nonstandard: it does not tend to $1$ but goes away from the reals when $\kappa^{\rm nor}$ tends to $0$, probably going to infinity in the lower complex half-plane.

To be complete and connect the case we study here with the one we will consider in Section III.C.2, or to a more general setting, let us mention that there are other \ns resonances which now tend to $0$ when $\nu$ tends to $0$. There is very likely an infinite number of them and they stem from the essential singularity at infinity in the complex plane of the oscillating \cf $G$ (behavior $e^{-i(i\infty)}$). Here are some of them, expressed in variable $\zeta$:
$$
\zeta=263-791\ i,\quad \zeta=461-814\ i,\quad \zeta=649-840\ i,\quad \zeta=829-866\ i,
$$

The resonance close to $1$ is more difficult to get exactly since it is very close to the reals, the integral in (7) being difficult to handle numerically. An estimation at first order in $\kappa^{\rm nor}$ gives
$$
\zeta= 1-9\ 10^{-6}-3.6\ 10^{-7}\ i.
$$  
An approached numerical calculation confirms the imaginary part. This yields a width of $1.4\ 10^{-5}$ eV. Let us note that the computation shows that when $\kappa^{\rm nor}$ grows beyond a critical value close to  $9.5$, it is this resonance which becomes a real negative eigenvalue, after having been zero at the critical point. The situation will be different in the case studied in the following section. If $\nu$ goes to $0$, this resonance tends to the above-defined  $\zeta_+$.

We already met these three types of behavior when $\nu$ goes to $0$, namely $\zeta(\nu)\rightarrow\zeta_-,\ \zeta(\nu)\rightarrow\zeta_+$ and $\zeta(\nu)\rightarrow 0$, in our preceding works. We will meet them again for the strong interaction at the Fermi scale that we are now going to consider. We are now going to get significant values. Let us anticipate and underline that the three types are a priori of equal importance from the theoretical point of view. Indeed, another way of presenting all the resonances is to classify them according to their behavior for $\kappa\rightarrow 0$; this will be done in Section 4.3 and we will then show that slightly changing the physical parameters of a system may make a resonance change from one class to another. This is related to the multivaluedness of the function the zero of which gives the resonances.

\smallskip

\centerline{\bf 2. ``Quarks'' and a strong coupling}

\smallskip

We take  $m=110$ MeV, $a=1$ Fermi and $\alpha=0.6$. We then get \lambdabar$\!\!_C/a=1.79$ and 
$$
E_{0,1}=1.74\  {\rm GeV},\quad E_{1,1}=3.55\  {\rm GeV},
$$
$$
g^{\rm nor}\simeq 0.45\ {\rm GeV},\quad \kappa^{\rm nor}=0.248,\quad\nu=0.11,\quad \chi^{\rm dist}=2.29.
$$
Numerical computations show that there are still two types of \ns resonances.

In the first class (first column of Table I which gives the corresponding complex energies) there is one resonance, at
$$
\zeta=0.03-0.11\ i.
$$
We put it in a first class because, when $\nu$ decreases, it approaches the reals, becomes zero (for $\nu\simeq 0.034$), and then real negative for $\nu<0.034$.
The resonance is \ns since one can show that it goes to infinity in the lower half-plane when  $\kappa^{\rm nor}$ tends to $0$. When  $\kappa^{\rm nor}$ increases, it approaches the reals,  becomes zero for a value close to $\kappa^{\rm nor}=0.44$, and then real negative (i.e. an eigenvalue) for $\kappa^{\rm nor}>0.44$.

In the second class (second column of the Table), we put an infinite family of resonances. They too seem to go to infinity in the lower half-plane when  $\kappa^{\rm nor}$ tends to $0$. But numerical computations show various limit points when $\kappa^{\rm nor}$ goes to infinity. Let us give three of these resonances:
$$
\zeta=0.81-0.43\ i\ ,\quad\zeta=1.22-0.63\ i ,\quad \zeta=1.59-0.75 \ i.
$$

Lastly, there is the \s resonance (third column), at 
$$
\zeta=1.082-7\ 10^{-5}\ i.
$$
Coming back to the energies, besides the eigenvalue  $E_{0,1}$ corresponding to state $f$ with no boson,  we get the following resonances

 \medskip

\setbox1\hbox{$1.80-0.21\ i$ GeV}
\setbox2\hbox{$\vcenter{\hbox{$3.21-0.79\ i$ GeV}\hbox{$3.95-1.14\ i$ GeV}\hbox{$4.62-1.36\ i$ GeV}\hbox{\quad\quad .......}}$}
\setbox3\hbox{$3.70-1.3\ 10^{-4}\ i$ GeV}

\boxit{\vbox spread 0.4cm{\vfil$$\hbox{\box1\quad\quad\quad\box2\quad\quad\box3}$$\vfil}}

\smallskip
\noindent
TABLE I. Energies of the resonances built with the first two levels $l=0$ and $l=1$ of a particle in a ball of radius $1$ F, coupled to the field ($\alpha=0.6$).

\medskip


The result of this section is that the coupling function's singularity at infinity yields interesting resonances in a case where $a/$\lambdabar$\!\!_C$ is of the order of unity. For the example\break $a/$\lambdabar$\!\!_C=0.56$, the singularity at infinity creates a resonance slightly above the fundamental, with a width of $210$ MeV, together with an infinite number of larger resonances; the real part of the first one sits between the two levels. In the $\nu\rightarrow 0$-limit, the two resonances of column 1 and 3 tend to the eigenvalues of the two eigenvectors of the Hamiltonian
$$
\big(E_{0,1}\dyad 00+E_{1,1}\dyad 11\big)\otimes 1+g^{\rm nor}\big(a^*(G)\ \otimes\dyad 01+a(G)\ \otimes\dyad 10\big),
$$
in the space spanned by states  $\ket 0\otimes\ket G$ and $\ket 1\otimes\ket{\rm vacuum}$. The others tend to the fundamental state's energy. (Note that here $\nu\rightarrow 0$ does not make the width of the continuum zero since it is originally infinite, despite the shape of the curve in Figure 1.)

Now that we saw that an appreciable value of $\chi^{\rm dist}$ can yield interesting \ns resonances, let us go back to the electron's case for two comments. Firstly, to get $\chi^{\rm dist}$ of the order of $1$ one should have to take dimensions much smaller than the Angstr\"om. But we do not see any physical meaning in taking $a$ of the order of 100 Fermi. Secondly, it is possible to find states $e$ and $f$ such that $\nu_{e,f}$ be close to $1$, i.e. such that the states' extent (in the ordinary sense we consider in this Section III) be of the order of the transition's wavelength. It suffices to take $p_2$ large for state $e$, say $p_2=15$. Then $\zeta_{1,p_2}=48.7$ and, with $a=5\Angstrom$, we have $\nu_{e,f}$ close to $1$.
But then the coupling strength decreases, and as a consequence $\chi_{e,f}^{\rm dist}$ remains small. This is due to the smallness of the electron's Compton wavelength, as formula (4) shows it. Moreover $||G||\simeq 0.11$; hence, with respect to the preceding $(p_2=1,\  a=1\Angstrom)$-case, $\chi^{\rm dist}$ becomes smaller by a factor of $150$. Therefore, having $\nu_{e,f}$ close to $1$ is not sufficient to ensure that \ns resonances are interesting: the coupling strength also has its importance.

We are now going to determine resonances in a case close to the preceding one but in which the states' extent is no longer constant: the extent will now depend on the state's excitation degree.

\bigskip

\centerline{\bf IV. THE RESONANCES WHEN THE PARTICLE} 

\centerline{\bf COUPLED TO THE FIELD IS IN A FINITE-DEPTH WELL.}

\smallskip
 In this case the \cf has singularities both at infinity and at a finite distance.
\medskip

\centerline{\bf A. Setting up and equation for the resonances associated to two levels}

\smallskip

We consider a one-dimension ($x$ variable ) square well, of depth $-V_0$ and width $a$, centered at the origin. The particle in this well has a mass $m$ and we will express the interesting quantities of the problem as functions of the dimensionless parameters\break $\alpha=$ \lambdabar$\!\!_C^{-1}\ a$ and $\beta=(mc^2)^{-1}\ V_0$. ($\alpha$ no longer denotes the coupling constants $\alpha_{\rm el}$ or $\alpha_{\rm s}$ of Section III.) For the ball, in the $a=1$ Fermi case and with the reduced mass of the same two quarks, these parameters took the values $\alpha=0.558$ and $\beta=\infty$.

For simplicity purpose, we will only consider even states. We recall the wave functions: 
$\Psi(x)=$\lambdabar$\!\!_C^{-1/2}\psi($\lambdabar$\!\!_C^{-1}x)$, of energies $E=\epsilon V_0$, with 
$$
\psi(\chi)=\left\{\matrix{N_1e^{\chi/\xi}\hfill&\quad{\rm if}\ \  \chi<-\alpha/2\hfill\cr N_1 sgn(u)(1+\epsilon)^{-1/2}e^{-\alpha/(2\xi)}\cos\big(\chi\sqrt{2\beta(1+\epsilon)}\big) &\quad{\rm if}\ \  -\alpha/2<\chi<\alpha/2\hfill\cr N_1e^{-\chi/\xi}\hfill&\quad{\rm if}\ \   \chi>\alpha/2\hfill\cr}\right.,
$$
$N_1=\Big(2\alpha^{-1} (1+\epsilon)\ e^{2\gamma\sqrt{-\epsilon}}\ \gamma\sqrt{-\epsilon}\ (1+\gamma\sqrt{-\epsilon})^{-1}\Big)^{1/2}$, $\gamma=\alpha\sqrt{{\beta\over 2}}$, $\xi=(-2\beta\epsilon)^{-1/2}$,\hfill\break $u=\cos\big(\alpha\sqrt{2^{-1}\beta(1+\epsilon)}\big)$, $sgn(u)$ is the sign of $u$, and $\epsilon$ is a solution for
$$
|\cos\big(\alpha\sqrt{2^{-1}\beta(1+\epsilon)}\big)|=\sqrt{1+\epsilon}\ ,\quad\tan\big(\alpha\sqrt{2^{-1}\beta(1+\epsilon)}\big)>0.
$$
Once we have chosen two eigenvectors $\ket 1$ and $\ket 2$ ($ E_1< E_2$) for the decoupled particle, we assume that the interaction of the particle with the field is described by a Hamiltonian $H_I$ whose matrix elements are
$$
g(k):=\obraket{1,k}{H_I}2=\lambda\int\Psi^*_1(x)e^{-ikx}\Psi_2(x)dx.\eqno (8)
$$
The aim in taking such an interaction Hamiltonian is to get a qualitative approach to the set of resonances one could find in more realistic interactions (see the sixth paragraph of the introduction.)
The calculus gives $g(k)=\lambda\  G\big((2\beta)^{-1/2}$\lambdabar$\!\!_C\ k\big)$, with
$$
G(q):=C_1{s_{12}\cos \gamma q-q\sin \gamma q\over q^2+s_{12}^2}+C_2\sum_{i,j=1,2}{\sin\Big(\gamma(y_1+(-1)^{i+1}y_2+(-1)^{j+1}q)\Big)\over y_1+(-1)^{i+1}y_2+(-1)^{j+1}q},             \eqno (9)
$$
$$
y_i=\sqrt{1+\epsilon_i},\quad  C_1=4s_1s_2y_1y_2C_2,\quad C_2={(\epsilon_1\epsilon_2)^{1/4}s_1s_2\over \sqrt{1+\gamma\sqrt{-\epsilon_1}}\sqrt{1+\gamma\sqrt{-\epsilon_2}}},
$$
$$
s_i=sgn[\cos\gamma y_i],\quad s_{12}=\sqrt{-\epsilon_1}+\sqrt{-\epsilon_2}.
$$
The normalized coupling constant is $||g||_2=(2\beta)^{1/4}$\lambdabar$\!\!_C^{-1/2}\ ||G||_2\  \lambda$; it has the dimension of an energy. The normalized \cf is 
$||g||_2^{-1}g$. ($||G||_2=\int_{-\infty}^\infty |G(q)|^2\ dq$). Restricting ourselves as usual to the Hilbert subspace spanned by $\ket 2$ and the $\ket{1,k}$'s, we get the equation for the eigenvalues:
$$
z-E_2-\int_{-\infty}^\infty {g(k)^2\over z-E_1-\hbar c|k|}dk=0.
$$
Through the change of variable $\zeta=z/V_0$ and introducing the dimensionless parameters
$$
\mu_1=\sqrt{2/\beta},\quad \kappa_1=(\hbar m c^3)^{-1/2} \lambda,  \eqno (10)
$$
we get the equation for the resonances:
$$
\zeta-\epsilon_2-\mu_1^3\kappa_1^2\int_0^\infty{G(q)^2\over \zeta-\epsilon_1-\mu_1 q}dq+2 i \pi \mu_1^2\kappa_1^2\ G(\mu_1^{-1}(\zeta-\epsilon_1))^2=0. \eqno (11)
$$
(Note that here $\mu_1$ is not the ratio of the transition wavelength to the (mean) extent of the two wave functions. Also $\kappa_1$ is not the ratio of the coupling strength to the levels' energy difference.) The mean extent, in the exponential sense from now on, of the two wave functions appears with the pole of $G$ at $q=-i\ s_{12}$. It corresponds to a pole of $g$ at $k=-i\  s_{12} $\lambdabar$\!\!_C^{-1}\sqrt{2\beta}$, the inverse of half the harmonic mean of the extents  \lambdabar$\!\!_C\xi_1$ and \lambdabar$\!\!_C\xi_2$ of the two states. This pole of $G$ and the singularity at infinity coming from the sine functions in (9) are at the origin of the \ns resonances.

We are now going to give the results of computer calculations of the resonances in some numerical examples. Our main interest is in examining the dependence of the resonances on the extent in space of the various level pairs.

We take $\mu_1=0.68$ and $\gamma=10$, corresponding to $\beta=4.32$ and $\alpha=6.8$. The particle in this well has four bound states. The reason of the choice for $\mu$ will appear in Sections 4.3 and 4.4. The energies of these bound states are (in  $V_0=4.32$ mc$^2$ units):
$$
\epsilon^{(1)}=-0.980,\quad \epsilon^{(2)}=-0.818,\quad\epsilon^{(3)}=-0.5,\quad\epsilon^{(4)}=-0.063.
$$
For the coupling to the field, we take $\kappa_1=1$, so as to have a strong coupling.

As for the ball's case, for each pair of levels, we have an infinity of resonances. To present them, we are going to classify them according to their behavior as $\kappa_1$ tends to $0$ (whereas the classification in Section III was based on the behavior with respect to $\nu$). Among the \ns resonances, two of them tend to the pole of $G$ at a finite distance, and an infinity of others tend to infinity. The diagram of Figure 5 gathers a large part of the results. Resonances are represented there by energy levels with a certain width.

\medskip

\centerline{\bf B. The only considered levels are levels 1 and 2. ($\epsilon_1=-0.980,\epsilon_2=-0.818$)} 

\smallskip

Let us first consider the pair of the first two levels. The mean extension , in  \lambdabar$\!\!_C$ units, is $s_{12}^{-1}\ \mu_1=0.36$. The coupling strength  $||g||_2$ is $(2\beta)^{1/4}\kappa_1||G||_2=0.91$, in mc$^2$ units. (We have $||G||_2=0.53$). In Figure 2 we give the graph of $||G||^{-1}|G|$, to which the graph of Figure 1 can be compared. We recall that $q=(2\beta)^{-1/2}$\lambdabar$\!\!_C\ k$.

\medskip
\setbox21=\hbox{\psboxto(5cm;0cm){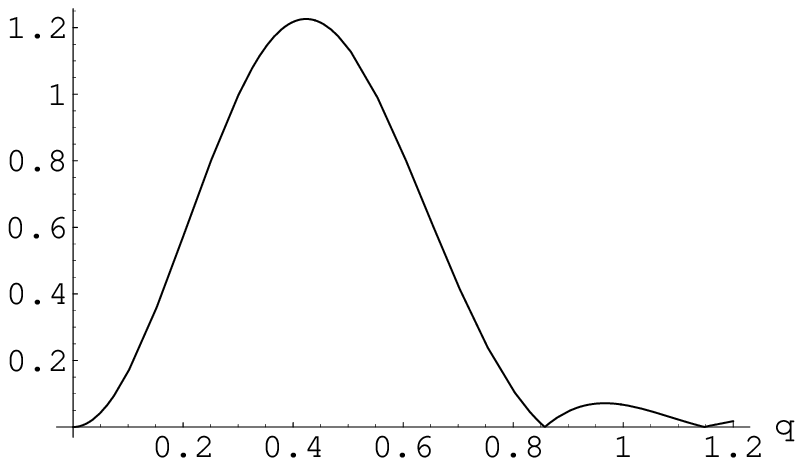}}
\setbox22=\hbox{\psboxto(5cm;0cm){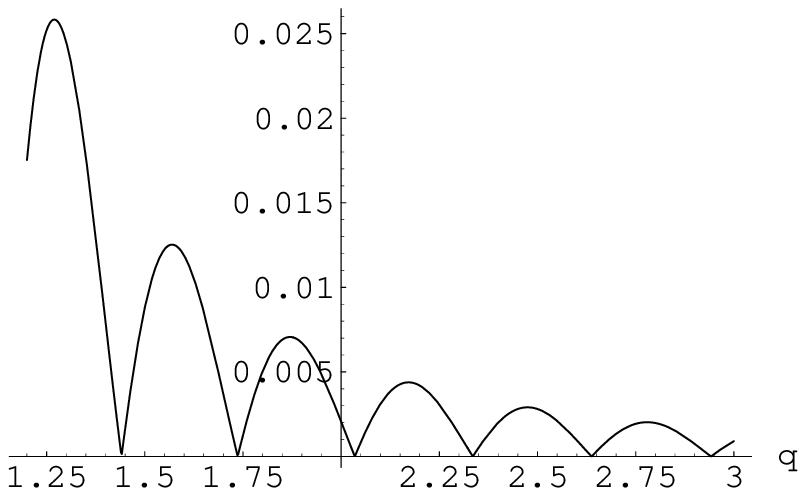}}
\hbox{\hskip 1cm\box21\hskip 2cm\box22}
\vskip 0.2cm
\centerline{FIG. 2. Graph of  $||G||^{-1}|G|$}

\medskip

i) The coupling to the field moves the upper state below the fundamental, at an energy
$$
\zeta=-0.983.
$$
Figure 3b shows the displacement of this resonance as $\kappa_1$ grows, starting from small values. It is a \s resonance.

ii) The two \ns resonances associated with the pole of $G$ at $-i\ s_{12}=-1.89 \ i$, which tend to $\epsilon_1-i\mu_1 s_{12}=\epsilon_1-1.29\ i$ when $\kappa_1$ goes to $0$, are
$$
\zeta=-0.491-0.0178\ i,\quad \zeta=-0.983-0.1\ i.
$$

iii) Lastly, there is an infinite family of resonances which go to infinity in the complex lower half-plane when $\kappa_1$ goes to $0$:
$$
-0.345-0.197\ i,\quad-0.138-0.278\ i,\quad +0.069-0.336\ i,\quad \cdots\ .
$$
We do not mention a resonance at $\zeta=-1.42-0.08\ i$; it is a replica of the one at $\zeta=-0.983-0.1\ i$ (see Section III.G).

\medskip

\centerline{\bf C. The only considered levels are levels 1 and 3. ($\epsilon_1=-0.980,\epsilon_2=-0.5$)}

\smallskip

Let us now consider levels 1 and 3. The mean extension is now $0.4$ and the coupling strength, $0.90$, is approximately the same as before.

i) The upper state is displaced to 
$$
\zeta=-0.264-5.8\ 10^{-3}\ i.
$$
There is a difference from the preceding case. It can be seen in Figure 3. The path which this (standard) resonance follows when $\kappa_1$ increases from $0$ to $1$ is given by curve (1) in Figure 3a. (Curve (2) is presented later on.) On Figure 3b we showed the displacement of the \s resonance for the level pair (1,2). (It passes below the fundamental for  $\kappa_1>0.98$.)

\smallskip

\setbox31=\hbox{\psannotate{\psboxto(5cm;0cm){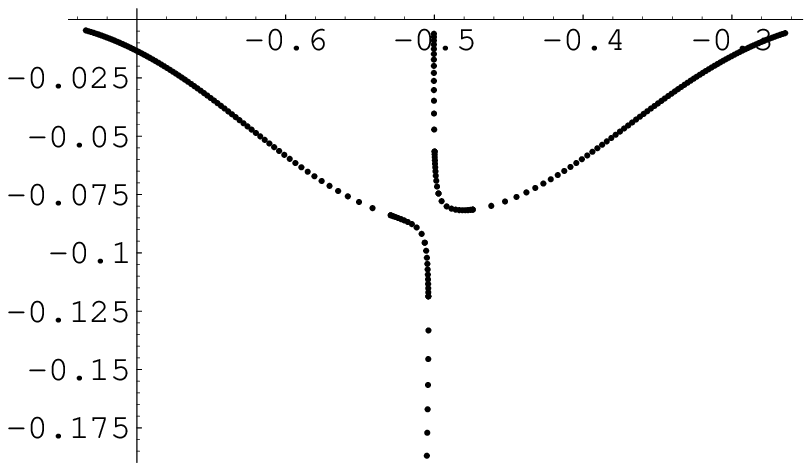}}{\at{3cm}{2cm}{\sevenrm(1)}\at{2.5cm}{-0.5cm}{(a)}\at{2cm}{1.5cm}{\sevenrm(2)}\at{3.5cm}{1.8cm}{$\nearrow\kappa_1$ {\sevenrm increases}}\at{2.8cm}{0.7cm}{$\downarrow\kappa_1$ {\sevenrm decreases}}\at{5cm}{2.7cm}{$\kappa_1=1$}}}
\setbox32=\hbox{\psannotate{\psboxto(5cm;0cm){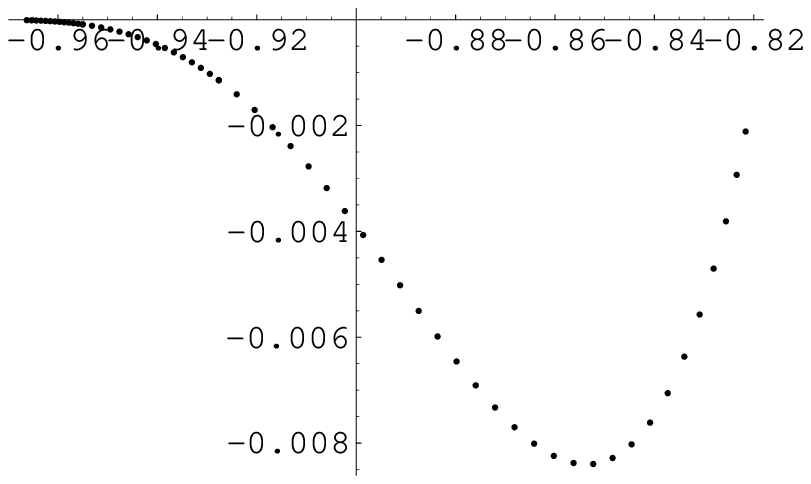}}{\at{2.5cm}{-0.5cm}{(b)}\at{5cm}{2.3cm}{$\nwarrow\kappa_1=0$}}}
\hbox{\box31\hskip 2cm\box32}
\vskip 0.6cm
\noindent
FIG. 3. Displacement of resonances with respect to the \cs for pairs 
(1,3) (a) and (1,2) (b).
\vskip 0.4cm

We will see later on that we can change the curve 1-type variation of the resonance into a Figure 3b-type through changing only slightly the well's geometry.

ii) The two resonances associated with the pole of $G$ at $-i\ s_{12}=-1.70 \ i$, which tend to $\epsilon_1-i\mu_1 s_{12}$ when $\kappa_1$ goes to $0$, are
$$
\zeta=-1.009-0.196 \ i,\quad \zeta=-0.830-0.162\ i.
$$

iii) Let us now come to the resonances which go to infinity when $\kappa_1$ tends to $0$. One of them is the one appearing on curve (2) of Figure 3a. Its value is 
$$
\zeta=-0.734-4.7\ 10^{-3}\ i.
$$
Although it is a \ns resonance, its width is remarkably small. We have just mentioned an explanation for this: the resonance can easily be changed into a \s one  by very little varying the physical parameters of the system. We develop this point in Section IV.D. It is therefore not surprising that its width be of the order of the widths of \s resonances. Others are for instance
$$
\zeta=-0.151-0.185 \ i\ ,\quad 0.060-0.268 \ i\ ,\quad \cdots\ . 
$$

\vfill\eject

\centerline{\bf D. Change of a \ns resonance into a \s one}
\centerline{\bf through a slight change in the  well's geometry}

\smallskip

Let us come back to the change of curve (1) in Figure 3a into a Figure 3b-type curve through a slight change in the  well's geometry. On Figure 3a we see that there are two resonances around  $\zeta\simeq-0.5-0.08 \ i$, for a value of $\kappa_1$ that a computer calculation puts near $0.25$. This computation shows that the relative position of curves  (1) and (2) in the neighborhood of the critical point changes from that of Figure 4a into that of Figure 4b as $\mu_1$ goes from $0.67$ to $0.69$.

\smallskip
\setbox41=\hbox{\psannotate{\psboxto(5cm;0cm){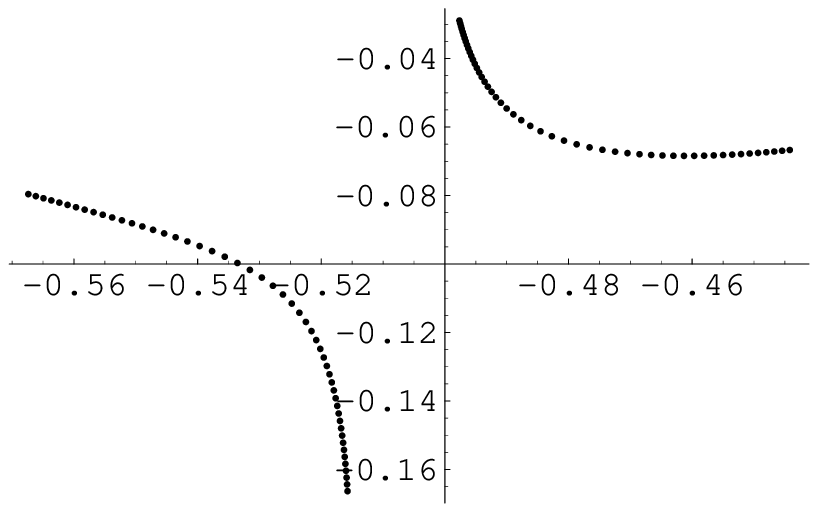}}{\at{2.5cm}{-0.5cm}{(a)}\at{3.5cm}{2.5cm}{\sevenrm(1)}\at{0.2cm}{0.3cm}{$\kappa_1$ {\sevenrm decreases}$\downarrow$}\at{1.3cm}{1cm}{\sevenrm(2)}\at{3.5cm}{1.9cm}{$\kappa_1$ {\sevenrm increases}$\rightarrow$}}}
\setbox42=\hbox{\psannotate{\psboxto(5cm;0cm){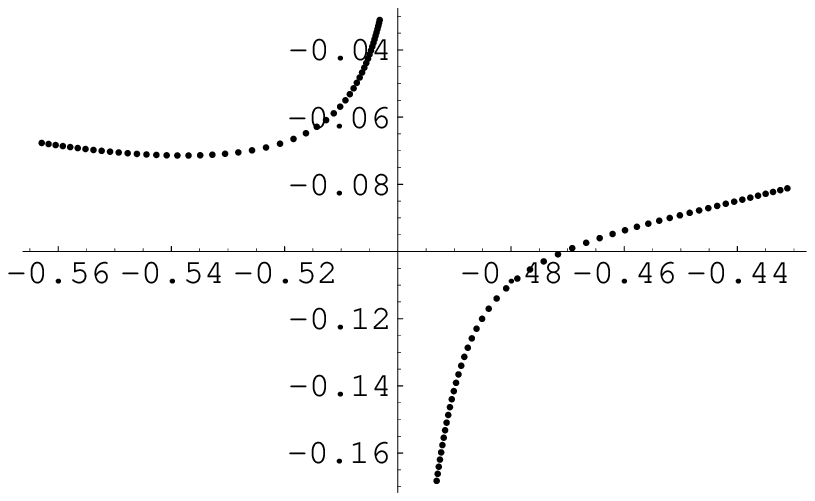}}{\at{2.5cm}{-0.5cm}{(b)}\at{1cm}{2.3cm}{\sevenrm(1)}\at{3cm}{0.7cm}{\sevenrm(2)}\at{0.2cm}{1.8cm}{$\leftarrow\kappa_1$ {\sevenrm increases}}\at{2.8cm}{0cm}{$\downarrow\kappa_1$ {\sevenrm decreases}}}}
\hbox{\box41\hskip 2cm\box42}
\vskip 0.6cm
\centerline{FIG. 4. Change of a \ns resonance into a \s one}

\vskip 0.5cm

\noindent
Thus the resonance at $-0.734-4.6\ 10^{-3}\ i$ for $\mu_1=0.68$, which sits on curve (2) of Figure 4a for  $\kappa=1$ (not shown on the figure) has its value slightly displaced for $\mu_1=0.69$, and now sits on curve (1) of Figure 4b. From a \ns resonance it has been changed into a \s resonance through the $\mu_1$-variation.

\medskip

\centerline{\bf E. Comparison of the results for pairs (1,2) and (1,3)}

\smallskip

The results concerning these two pairs of levels, with coupling strengths to the field nearly identical, lead to the following comments.

Firstly, it follows from Sections IV.C and IV.D that one cannot deduce the order of magnitude of the imaginary part of a resonance from its \s or \ns type, since this type may change under a small variation of the parameters. All the resonances are to be put on the same footing, a priori, even those coming from infinity since we get here an example of a small width resonance although it comes from infinity, for small $\kappa_1$.

The instability of the three classes complicates the study of the effect of the states' extent on the resonances' position, since the extent is clearly related to one class only, the (ii) one, connected to the poles of $G$.

Nevertheless we can say the following. The pole at a finite distance (in the lower half-plane) of the \cf approaches the reals when we switch from pair (1,2) to pair (1,3), since the mean extent increases. However, we notice that the imaginary part of the resonance coming from this pole increases, from one pair to the other.
Therefore, the position of these poles does not give a precise indication about the position of the resonances which are attached to them, besides the case where the coupling would be small enough for the resonance to be close to the pole. Let us note that this may be a consequence of the existence of singularities at infinity.

Still as regards the resonances coming from the pole at a finite distance, we nevertheless note that the real part of the energy of the lower resonance decreases as the spatial extent of the pair increases.
The numerical values for pair (1,4) will confirm these points.

\medskip

\centerline{\bf F. The only considered levels are levels 1 and 4 ($\epsilon_1=-0.980,\epsilon_2=-0.063$)}

\smallskip
The mean extent is now  $0.55$ and the coupling strength, $0.81$, has decreased.

i) The upper state's energy is displaced at 
$$
\zeta=+0.04-1.5\ 10^{-3}\ i.
$$
a displacement which goes in the same direction as the one found for pair (1,3).

ii) The two nonstandard resonances associated with  the pole of $G$ at $-i\ s_{12}=-1.24 \ i$, that is tending to $\epsilon_1-i\mu_1\ s_{12}=-0.980-0.844\ i$ when $\kappa_1$ tends to $0$, are
$$
\zeta=-1.013-0.25 \ i,\quad \zeta=-0.85-0.24\ i.
$$
Here also the lower resonance's energy has decreased with respect to pair (1,3)'s case. We still notice that the imaginary part is greater although the coupling is weaker and the pole of $G$ closer to the reals. Also, with respect to the energy of the upper uncoupled state, $-0.818$ for pair $(1,2)$ and $-0.063$ for pair $(1,4)$, we note that the upper resonance is much lower for pair $(1,4)$: $-0.85$, than for  pair $(1,2)$: $-0.49$, even if we take account of the width. These numbers show that the pole of $G$ at a finite distance does not play a role through its imaginary part only: the real part ($e_1$ in the three cases) of the pole it induces in integral (7), pole which is responsible for the existence of these resonances, is also important.

iii) Lastly, as regards the resonances coming from infinity at small coupling, there is still an infinity of them. One the one hand,
$$
\zeta=-0.48-0.045\ i,\quad \zeta=+0.45-0.32\ i,\quad\cdots,
$$
one of them having a quite small imaginary part. On the other hand 
$$
\cdots\ ,\quad\zeta=-2.20-0.324\ i,\quad\cdots,\quad \zeta=-1.99-0.264\ i, 
$$ 
which come within the discussion in paragraphs (b) and (c) of Section IV.H.

Figure 5 gathers the energies of some of the resonances we obtained, with their widths, in $V_0$ units. It shows that the real part of the resonances coming from the pole of $G$ at a finite distance decreases as the space extent increases. But the imaginary part increases, contrary to what might have been expected. But comparing pair (1,3) and (2,3), we are now going to see that a somewhat larger extent do lead to a smaller width.

\vfill\eject

$$
\psannotate{\psboxto (0cm;9cm){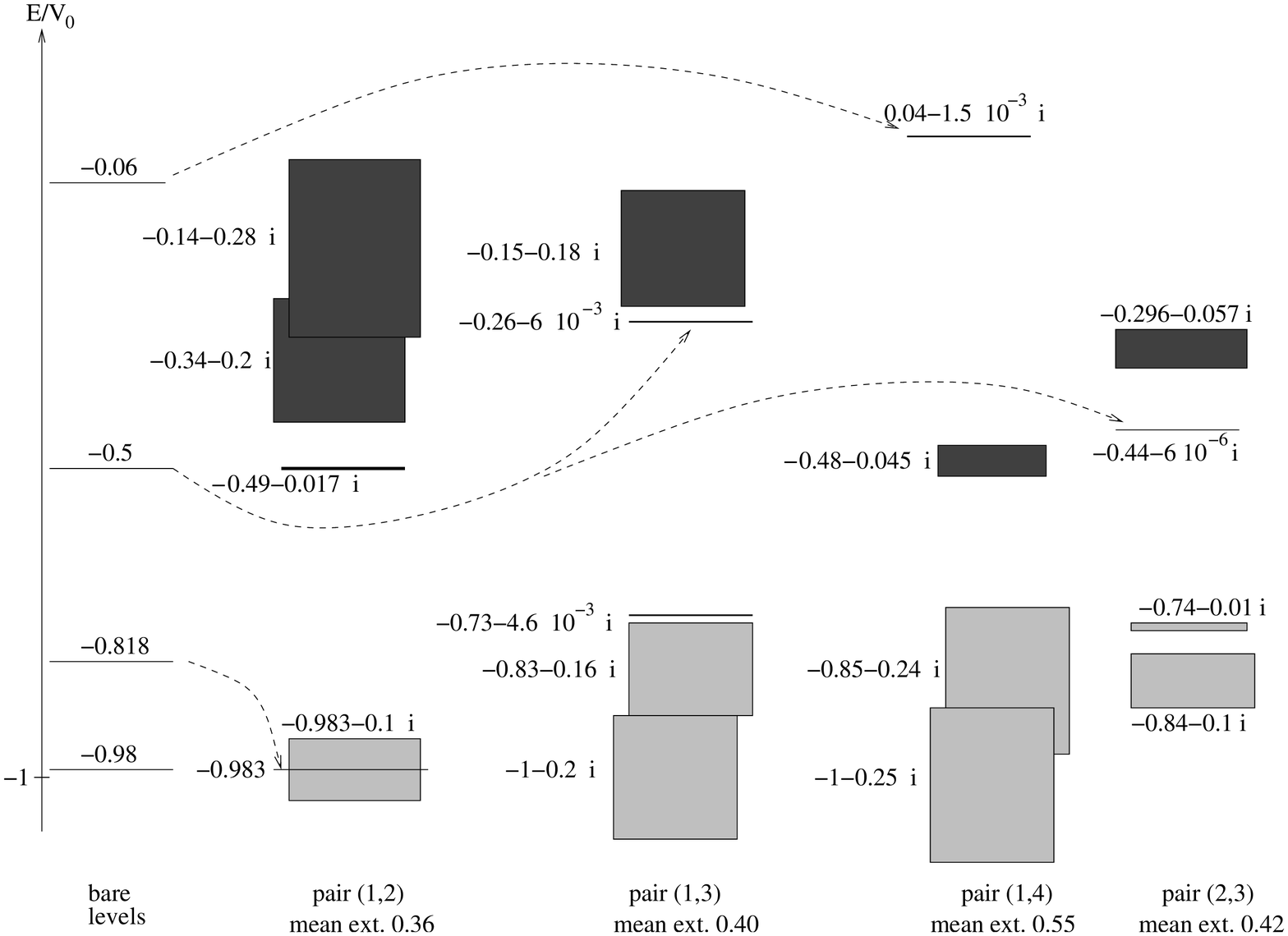}}{\at{1.8cm}{4.6cm}{\sevenrm from the pole at finite dist.}\at{5.2cm}{3.3cm}{\sevenrm from sing. at infinity}}
$$ 
FIG. 5. Representation of some resonances with their widths, for various level pairs. The pairs' mean extent is indicated  in \lambdabar$\!\!_C$ units.
Resonances represented in dark gray, together with the resonance at $\zeta= -0.73-4.6\ 10^{-3}\ i$, are those which come from (i.e. whose $\lambda\rightarrow 0$)-limit is) the singularity of $g$ at infinity. Those in light gray, together with the one at $-0.49-0.02\ i$, come from the poles at a finite distance. We connected the standard resonances to the corresponding bare levels' energies by a dashed line. For each pair, the level  $e_1=-0.98$, which remains an eigenvalue has not been drawn.


\medskip
\centerline{\bf G. The only considered levels are levels 2 and 3. ($e_1=-0.818,\ e_2=-0.5$)}
\smallskip 

The mean extent is  $0.42$ \lambdabar$\!\!_C$ and the \cs $0.9$  mc$^2$.

i) The upper state is displaced by the coupling to the field at 
$$
\zeta=-0.437-5.7\ 10^{-6} i.
$$

ii) The two \ns resonances associated to the pole of $G$ at $-i\ s_{12}=-1.61 \ i$, which tend to $\epsilon_1-i\mu s_{12}=\epsilon_1-1.1\ i$ when $\kappa_1$ tends to $0$, are
$$
\zeta=-0.84-9.84\ 10^{-2}i, \quad \zeta=-0.74-9.9\ 10^{-3}i.
$$
As for the three preceding cases, the real part of the lower one passes below the lower bare level of the pair. But now its imaginary part, i.e. the width, is smaller than the widths for the other pairs.

iii) Resonances which are likely to go to infinity for  $\kappa_1\rightarrow 0$ are 
$$
\zeta=-0.296-0.057\ i,\quad\zeta=+0.032-0.064\ i,\cdots\ .
$$
Note that the widths are quite small, as for (ii). Some of the resonances whose real parts are negative are represented in the last column of Figure 5.

There is also
$$
\cdots\ ,\quad\zeta=-1.35-0.101\ i,\quad\zeta=-1.135-0.062\ i.
$$
These last values, together with the last resonances of paragraph (iii) of Section IV.F lead us to the following discussion.

\medskip

\centerline{\bf H. DISCUSSION} 

\smallskip

For each pair, we find a great number of resonances, probably because of the infinite order ``pole'' at infinity. One may ask the question whether some of them could be rejected as having no physical signification. It is an involved problem. Here are some observations. 

a) The first rejection criterion may be the magnitude of the imaginary parts. The physical importance of the resonances appears in the behaviors in $t$ of  $\obraket{e_1,k}{\exp{-iHt/\hbar}}{e_2}$ or of $U_2(t):=\obraket{e_2}{\exp{-iHt/\hbar}}{e_2}$. The resonances also manifest themselves in peaks of the function  $k\rightarrow \displaystyle\lim_{t\rightarrow\infty}\obraket{e_1,k}{\exp{-iHt/\hbar}}{e_2}$.  $U_2(t)$ is computed through a complex integration which has two contributions. The first one comes from the poles of $\obraket{e_2}{[z-H]^{-1}}{e_2}$, analytically continued into the complex lower half-plane across $\bbbr^+$. They are the resonances we determined. If their imaginary parts are too large, the contributions of these poles yield fast exponential decreases of $U_2(t)$ or indiscernible peaks in the graph of function $k\rightarrow \displaystyle\lim_{t\rightarrow\infty}\obraket{e_1,k}{\exp{-iHt/\hbar}}{e_2}$. Although the order of magnitude of the widths of the resonances we encountered are various, we do not see any reason to put some resonances aside according to this criterion, except for the widest in the infinite series.

b) A second criterion is the magnitude of the residues of $\obraket{e_2}{[z-H]^{-1}}{e_2}$ at its poles. We will not enter into this discussion, partly because of the following point. The second contribution of the above-mentioned complex integration comes from a cut starting at the origin, a cut which we have to chose in the lower half-plane (see [8], A$_{III}$.4). Its position determines which resonances are to be taken into account. As a consequence, one should be able to calculate the contributions of the various possible cuts in order to know which resonances play a role. We did not address this aspect of our problem. We limited ourselves to presenting the structure of the whole set of resonances, focusing on the point that the \s resonance, \ns resonances coming from a pole at a finite distance and \ns resonances coming from a singularity at infinity are a priori to be put on the same footing. As regards the two first types, a convincing example was already [2] the Rabi doublet in the cavity case.  As regards resonances coming from a singularity at infinity, Section IV.C gives a physical argument to put one of them on the same footing as a \s resonance.

c) Even if the choice of the cut is not taken into account, one has however to note that one resonance may be a kind of replica of another one. Let us show this.
In [1] we considered a two level model, the coupling to the field being given by the family of functions $g_\nu=\nu^{-1/2}g(\nu^{-1}p)$, $\nu$ being a parameter which dilates $g(p)=\sqrt{2/\pi}\ p/(1+p^2)$. We showed that if $\nu$ is smaller than  $\nu_c(\lambda)=2\int_0^\infty p^{-1}\ g(p)^2dp\ \lambda^2\simeq 0.637\ \lambda^2$ then the Hamiltonian has a real negative eigenvalue $\zeta(\nu)$, which vanishes if $\nu=\nu_c(\lambda)$. Let  $\zeta_-$ denotes the eigenvalue for $\nu=0$. When $\nu$ increases past $\nu_c$, the eigenvalue changes into a resonance in the complex lower half-plane, a zero of the analytic continuation across the cut $\bbbr^+$ of the function 
$$
z\rightarrow f(\nu,z)=z-1-2\lambda^2\int_0^\infty{g(p)^2\over z-\nu p}\ dp.
$$ 
(The energies of the two levels are respectively $0$ and $1$.) For small $\nu$, $\zeta(\nu)$ is close to $\zeta_-$. It is a zero of $f(\nu,.)$. But, still for small $\nu$, one can show that the analytic continuation of $f(\nu,.)$, which is denoted by $f_+(\nu,.)$, also has a zero near $\zeta_-$, which is  complex. Let us denote it by $\zeta_+(\nu)$. If we consider the zeros of the branches $f_+,f_{++},\cdots$ obtained through turning clockwise once, twice, etc.., round the branch point $0$, we get an infinity of points. $\zeta_+(\nu,.)$ is a kind of replica of $\zeta(\nu,.)$ and cannot be considered as a new resonance. As a consequence, out of all these zeros, only $\zeta(\nu)$ will be kept in the list of relevant resonances. It may be noted that the above-mentioned cut could be chosen in such a way that the residue at $\zeta_+(\nu,.)$ contributes; however there are other choices for which $\zeta_+(\nu)$ does not play any role.

To sum up the discussion of this section we could say that the paper presents resonances which are important but that it is not possible at this stage to give a complete list of all those which have a physical meaning.

However, it is perhaps not worth carrying the study too far in the case of the well since the complication coming from the singularity of the \cf at infinity would perhaps not occur in realistic potentials for strong interactions. We already said that the complication is not present for the $1/r$ potential. (We would nevertheless have to cope with the multiplicities of the poles.)

\bigskip

\centerline{\bf V. CONCLUSION}

\smallskip

Let us recall that in the whole study we had to select two levels of the uncoupled system $S$, an excited one ($e_2$), and the other one ($e_1$) considered as the fundamental state; we forgot all the others. Due to this bias our results can only be qualitative ones. But we do not have any equation for the resonances if there are more than two levels. Besides, we used models and do not at all pretend that we got a quantitative approach to a description of resonances in strong interaction physics.

We showed that the resonances, poles of the continuation of $\obraket{e_2}{[z-H]^{-1}}{e_2}$ in the complex lower half-plane of the $z$ variable, are in principle as important as the eigenvalues of $H$, and that all of them are to be considered. Indeed, the imaginary part of a resonance which disappears when the coupling vanishes may sometimes be so small that the resonance is clearly to be taken into account physically (Section IV.C).
In the hydrogen atom case, we had shown [2] that the ratio of the widths of the \s resonances to those of the \ns resonances is of the order of $10^{-11}$, very small as a consequence of the smallness of both the coupling constant and the Compton wavelength of the electron. For the electron in a ball of radius $1$ Angstr\"om, coupled to the transverse electromagnetic field, we saw here that this ratio is of the same order,  $3\ 10^{-10}$. To push the \ns resonances closer to the reals, one should diminish $a$. But does it make sense to consider an electron confined inside a ball whose radius is smaller than its Compton wave length? Considering more massive particles (for example $\sim 100$ MeV), with a ball's radius of $1$ Fermi and a strong coupling to the field, we showed that the ratio lowers to values of the order of $3\ 10^{-4}$. We then get widths of a few hundred MeV, comparable to those found experimentally in strong interaction physics, although they are still large with respect to the width of the excited state.

We were able to diminish this ratio still more drastically through considering a particle in a finite depth well, with a strong coupling to the field. Indeed we showed a case where this ratio can be nearly equal to $1$ (Section IV.C). We also showed how we can make \ns resonances narrower through choosing more extended states (Section IV.G). This is due to the following fact: in a perturbative approach, that is if one start with the uncoupled system S, the coupling (not necessarily small) of a two-level system with a field is described by a function which depends on the two levels and whose analytic structure depends strongly on the space extent of the wave functions of these levels. Nevertheless, we saw that this spreading-induced narrowing is not automatic. It would be interesting to test it with a greater number of levels or when singularities at infinity are absent since they may perturb the connection between the imaginary parts of the poles of the coupling function and the resonances' widths. In strong interactions less elementary than the ones we considered here the set of resonances could be simpler.

We also found resonances which are narrow for other reasons than the excitation degree of the levels. This is due to the instability of the splitting of the set of resonances into \s and \ns resonances (Section IV.C and IV.D). One thus sees the usefulness of considering all the resonances.

\bigskip

\centerline{\bf APPENDIX: EQUATION FOR THE EIGENVALUES,} 
\centerline{\bf FOR A PARTICLE COUPLED TO A MASSLESS VECTOR FIELD}

\smallskip

We assume that only two states of the particle come into play: an excited state, with wave function $\psi_{l_2,m_2,p_2}(\vec r)$ and a fundamental one with wave function $\psi_{l_1,m_1,p_1}(\vec r)$, toward which the former can go through emitting a massless vector boson with state function $\vec s(\vec k)$. Couples $(l_2,m_2),(l_1,m_1)$ of the particle's angular momenta are assumed to be fixed. We shorten $\psi_{l_i,m_i,p_i}$ to $\psi_i$.

The interaction Hamiltonian is 
$$
H_I={-i\hbar^2\over 2\pi m}\alpha^{1/2}\sum_{j=1,2}\int_{-\infty}^\infty{1\over\sqrt{k}}\big(c_j^*(\vec k)e^{-i\vec k\cdot\vec r}+c_j(\vec k)e^{i\vec k\cdot\vec r}\big)\vec\epsilon_j(\vec k)\cdot\vec\nabla\ d\vec k.  \eqno ({\rm A1})
$$
Through using the spherical harmonics (see page 18 of [8]):
$$
\vec Y_{jm}^{(e)}(\hat k)={1\over \sqrt{j(j+1)}}\vec\nabla_{\hat k}Y_{jm}(\hat k),\quad 
\vec Y_{jm}^{(m)}(\hat k)=\hat k\times Y_{jm}^{(e)}(\hat k),\quad \vec Y_{jm}^\lambda(\hat k)={i\vec Y_{jm}^e(\hat k)-\lambda\vec Y_{jm}^m(\hat k)\over\sqrt{2}},
$$
the states of a boson with angular momenta $(j,m)$ and helicity $\lambda$ are described by $\vec s_{j,m,\lambda}(\vec k)$\break$=s(k)\vec Y_{jm}^\lambda(\hat k)$. (The scalar product in the boson-state space is $\int\bar s_1(\vec k)s_2(\vec k)\ d\vec k$.) In these formulas, $\vec\nabla_{\hat k}$ is $k$ times the projection of the gradient $\vec\nabla_{\vec k}$ on $\vec k^\perp$, the subspace orthogonal to $\vec k$. Due to our initial assumption, the only possible states of the electron-boson system are in the subspace ${\cal E}_1$ spanned by the $\psi_1\otimes \vec s$ 's and $\psi_2\otimes\Omega$, where $\vec s$ is a linear combination of the $\vec s_{j,m,\lambda}$'s  and $\Omega$ denotes the vacuum of the radiation. Let $P_{{\cal E}_1}$ denotes the orthogonal projection on ${\cal E}_1$. We reduce the original Hamiltonian to $H_I^{\rm app}:=P_{{\cal E}_1}H_IP_{{\cal E}_1}$, which acts in ${\cal E}_1$; it has the same matrix elements as $H_I$. So we neglect matrix elements between states like  $\psi_1\otimes \vec s$ and $\psi_{\alpha_3,l_3,m_3}$, or between $\psi_2\otimes \vec s$ and $\psi_1$. We also forget contributions from other angular momenta.

Set   $\vec F_{12}(\vec k):=\int \bar \psi_1(\vec r)\ e^{-i\vec k\cdot\vec r}\ \vec\nabla_{\vec r}\psi_2(\vec r)\ d\vec r$. With $C:={-i\hbar^2\over 2\pi m}\alpha^{1/2}$, and $P_{k^\perp}$ denoting the projector on $\vec k^\perp$, we have
$$
H_I^{\rm app}(\psi_2\otimes\Omega)=\psi_1\otimes\{\vec k\rightarrow C k^{-1/2}P_{\vec k^\perp}\vec F_{12}(\vec k)  \},     \eqno ({\rm A2a})
$$
$$
H_I^{\rm app}(\psi_1\otimes\vec s)=-C\ \Bigg(\int_{\bbbr^3}k^{-1/2}\vec s(\vec k)\cdot (P_{\vec k^\perp} \bar{\vec F_{12}}(\vec k))\Bigg) \psi_2\otimes \Omega.\eqno (A2b)
$$
Looking for an eigenvector in ${\cal E}_1$ of the form $\psi_1\otimes\vec s+\psi_2\otimes\Omega$, with $z$ its eigenvalue, we find 
$$
\vec s(\vec k)=-C\ k^{-1/2}(E_1-z+\hbar c k)^{-1} P_{\vec k^\perp}\vec F_{12}(\vec k),                                                       \eqno (A3)
$$
with 
$$
z-E_2-C^2\int_{\bbbr^3} k^{-1}(E_1-z+\hbar c k)^{-1} |P_{\vec k^\perp}\vec F_{12}(\vec k)|^2\ d\vec k=0,                                  \eqno (A4)
$$
which is the equation we looked for. From now on we assume that $l_1=0$. Introducing the radial parts of the wave functions, we have 
$\psi_1(\vec r)=(4\pi)^{-1/2} R_{0,p_1}(r)$ and $\psi_2(\vec r)=R_{l_2,p_2}(r)Y_{l_2m_2}(\Omega_{\hat r})$. Setting 
$$
I_ {p_1,l_2,p_2}(k)=\int_0^\infty r R'_{0,p_1}(r)R_{l_2,p_2}(r)j_{l_2}(kr)\ dr,\eqno (A5)
$$
we get
$$\displaylines{\eqalign{
P_{\vec k^\perp}\vec F_{12}(\vec k)&=i^{l_2-1}\sqrt{4\pi}\  I_{p_1,l_2,p_2}(k)\big(P_{\vec k^\perp}\vec\nabla_{\vec k}Y_{l_2m_2}(\hat k)\big)\cr 
                                 &=2\sqrt {l_2(l_2+1)\pi}\ i^{l_2-1}\ k^{-1}\  I_{p_1,l_2,p_2}(k)\ \vec Y_{l_2m_2}^{(e)}(\hat k).  
}}$$
Through using $\int\Big(\vec Y_{lm}^{(e)}(\hat k)\Big)^2 d\Omega_{\hat k}=1$, we get the eigenvalue equation
$$
z-E_2-l_2(l_2+1){\hbar^4\alpha\over \pi m^2}\int_0^\infty k^{-1} { I_{p_1,l_2,p_2}(k)^2\over z-E_1-\hbar c k}\ dk=0.                                       \eqno (A6)
$$
Noticing that $I_{p_1,l_2,p_2}(k)=a^{-2}G_{p_1,l_2,p_2}(ka)$, we obtain formulas (2) and (3) of the text, with $y=ka$.

\vfill\eject


\bigskip

\item{[1]} C. Billionnet,  {\it J. Phys.} A {\bf 35}, 2649 (2002)

\smallskip
\item{[2]} C. Billionnet, {\it Eur. Phys. J.} D {\bf 41}, 9 (2007)

\smallskip
\item{[3]} C. Billionnet, {\it Int. J. Mod. Phys.} {\bf 19}, 2643 (2004)

\smallskip
\item{[4]} C. Billionnet, {\it Eur. Phys. J.} D {\bf 8}, 157 (2000)

\smallskip
\item{[5]} C. Billionnet, {\it J. Math. Phys.} {\bf 46}, 072101 (2005)

\smallskip
\item{[6]}   H. E. Moses {\it Phys. Rev.} A {\bf 8}, 1710 (1973)

\smallskip
\item{[7]} C. Cohen-Tannoudji, J. Dupont-Roc and G. Grynberg, {\it Processus d'interaction entre photons et atomes} (CNRS, Paris, 1988) (Engl. transl. Wiley, New York, 1992)

\smallskip
\item{[8]} L. D. Landau and E. M. Lifshitz, {\it Course of Theoretical Physics}, Vol. 4, Part 1 (Pergamon Press, 1971)

\bye
\end